\documentclass[12pt]{article}

 \newcommand{\ben}{\begin{enumerate}}
\newcommand{\een}{\end{enumerate}}
\newcommand{\beq}{\begin{equation}}
\newcommand{\eeq}{\end{equation}}
\newcommand{\bse}{\begin{subequation}}
\newcommand{\ese}{\end{subequation}}
\newcommand{\bea}{\begin{eqnarray}}
\newcommand{\eea}{\end{eqnarray}}
\newcommand{\bc}{\begin{center}}
\newcommand{\ec}{\end{center}}

\def\DR{\rm I\kern-1.45pt\rm R}
\def\DC{\kern2pt {\hbox{\sqi I}}\kern-4.2pt\rm C}
\def\DH{\rm I\kern-1.5pt\rm H\kern-1.5pt\rm I}
\begin{document}

\begin{center}
{\Large\bf The generalized MIC-Kepler system}\\[3mm]
{\bf Levon Mardoyan} \\[3mm]
International Center for Advanced Studies,\\ Yerevan State
University, \\
1, Alex Manoogian st., 375025, Yerevan, Armenia
\end{center}

\vspace{3mm}
\begin{abstract}
This paper deals with the dynamical system that generalizes the
MIC-Kepler system. It is shown that the Schr\"{o}dinger equation
for this generalized MIC-Kepler system can be separated in
spherical and parabolic coordinates. The spectral problem in
spherical and parabolic coordinates is solved.
\end{abstract}

\vspace{0.5cm}

\setcounter{equation}{0}

\section{Introduction}

The system described by the Hamiltonian
%================================================================
\bea \hat{\cal H}=\frac{1}{2}(-i{\bf{\nabla}}- s{\bf A})^2
+\frac{{s}^2}{2 r^2}-\frac{1}{r}+\frac{c_1}{r(r+z)} +
\frac{c_2}{r(r-z)}, \label{1} \eea
%==================================================================
where $c_{1}$ and $c_{2}$ nonnegative constants, later on will be
called the generalized MIC-Kepler system. (We use the system of
units for which $\hbar=m=e=c=1$.)

The MIC-Kepler integrable system was constructed by Zwanziger
\cite{Z} and rediscovered by McIntosh and Cisneros \cite{mic}.
This system is described by the Hamiltonian
%===================================================================
\begin{equation}
\hat{{\cal H}_0}=\frac{1}{2}(-i{\bf{\nabla}}- s{\bf A})^2
+\frac{{s}^2}{2 r^2}-\frac{1}{r},\label{2}
\end{equation}
%====================================================================
where
%====================================================================
\begin{eqnarray*}
{\bf A}=\frac{1}{r(r-z)}(y,-x,0), \qquad {\rm and} \qquad {\rm
rot}{\bf A}=\frac{{\bf r}}{r^3}.
\end{eqnarray*}
%=====================================================================

Its distinctive peculiarity is the Coulomb  hidden symmetry given
by the following constants of motion (\ref{2})
%=====================================================================
\begin{equation}
\hat{\bf I}=\frac{1}{2}\left[(-i{\bf\nabla}-s{\bf
A})\times{\hat{\bf J}}-\hat{\bf J}\times (-i{\bf\nabla}-s{\bf
A})\right] +\frac{{\bf r}}{r},\quad \hat{\bf J}={\bf r}\times
(-i{\bf\nabla}-s{\bf A})
 - s\frac{{\bf r}}{r}.
\label{3}
\end{equation}
%======================================================================
Here, the operator $\hat{\bf J}$ defines the angular momentum of
the system, while operator $\hat{\bf I}$ is the analog of the
Runge-Lenz vector. These constants of motion, together with the
Hamiltonian, form the quadratic symmetry algebra of the Coulomb
problem. For fixed negative energy values the motion integrals
make up algebra $so(4)$, whereas for positive energy values -
$so(3.1)$. Due to the hidden symmetry the MIC-Kepler problem is
factorized not only in the spherical but parabolic coordinates as
well. Hence, the MIC-Kepler system is a natural generalization of
the Coulomb problem in the presence of Dirac's monopole. In both
cases the monopole number $s$ satisfies the Dirac's rule of charge
quantization $s=0,\pm1/2,\pm 1,\ldots$.

The MIC-Kepler system could be constructed by the reduction of the
four-dimensional isotropic oscillator by the use of the so-called
Kustaanheimo-Stiefel transformation both on classical and quantum
mechanical levels \cite{nt}. In the similar way, reducing the two-
and eight- dimensional isotropic oscillator, one can obtain the
two-  \cite{ntt}  and five-dimensional  \cite{iwa1} analogs of
MIC-Kepler system. An infinitely thin solenoid providing the
system by the spin $1/2$, plays the role of monopole in
two-dimensional case, whereas in the five-dimensional case this
role is performed by the $SU(2)$ Yang monopole \cite{yang},
endowing the system by the isospin. All the above-mentioned
systems have Coulomb symmetries and are solved in spherical and
parabolic coordinates both in discrete and continuous parts of
energy spectra \cite{mardoyan}. There are generalizations of
MIC-Kepler systems on three-dimensional sphere \cite{kurochkin}
and hyperboloid  \cite{np} as well. The MIC-Kepler system has been
worked out from different points of view in Refs.~\cite{mladenov,
iwa2, junker, bn, mps}.

For integer values $s$ the MIC-Kepler system describes the
relative motion of the two Dirac's dyons (charged magnetic
monopoles), where vector ${\bf r}$ determines the position of the
second dyon with respect to the first one \cite{Z}. For
half-integer $s$ the presence of the solenoid magnetic field,
endowing the system with the spin $1/2$, is presupposed (see, e.g.
\cite{ntt}).

The Hamiltonian (\ref{1}) for $s=0$ and $c_{i}\neq 0$ $(i=1,2)$
reduces to the Hamiltonian
%===============================================================
\bea \hat H=-\frac{1}{2}\Delta -\frac{1}{r}+\frac{c_1}{r(r+z)} +
\frac{c_2}{r(r-z)}, \label{4} \eea
%===============================================================
of the generalized Kepler-Coulomb system \cite{KMP}.

The potential
%================================================================
\bea
V = -\frac{\alpha}{r}+\frac{c_1}{r(r+z)} +
\frac{c_2}{r(r-z)}, \label{5}
\eea
%===============================================================
is one of the Smorodinsky-Winternitz type potentials \cite{sw}.
The Smorodinsky-Winternitz type potentials where revived and
investigated in the 1990 by Evans \cite{ev}. In the case where
$c_{1}=c_{2}$, the potential (\ref{5}) reduces to the Hartmann
potential that has been used for describing axially symmetric
systems like ring-shaped molecules \cite{hart} and investigated
from different points of view in Refs. \cite{kn}-\cite{zhed}. In
particular, the (quantum mechanical) discrete spectrum for the for
the generalized Kepler-Coulomb system (\ref{4}) is well known
\cite{guha, cber, drag}, even for the so-called $(q,p)$-analogue
of this system \cite{drag}. Furthermore, a path integral treatment
of the potential (\ref{5}) has been given in Refs.~\cite{hamm2,
cber}. Recently, the dynamical symmetry of the generalized
Kepler-Coulomb system has been studied in Refs.~\cite{drag, kicam,
zhed}, the classical motion of a particle moving in the potential
(\ref{5}) has been considered in \cite{kicam}, and the
coefficients connecting the parabolic and spherical bases have
been identified in \cite{zhed} as Clebsch-Gordan coefficients of
the pseudo-unitary group $SU(1,1)$.

The purpose of the present paper is to further study the bound
states of the generalized MIC-Kepler system in spherical and
parabolic coordinates.

\setcounter{equation}{0}

\section{Spherical Basis}

The Schr\"{o}dinger equation with Hamiltonian (\ref{1}) in
spherical coordinates $(r, \theta, \varphi)$ may be solved by
seeking a wavefunction $\psi$ of the form
%=================================================================
\bea
\psi(r,\theta,\varphi)=R(r)Z(\theta, \varphi). \label{6}
\eea
%=================================================================
This amounts to finding the eigenfunctions of the set $\{\hat{\cal
H}, \hat{J_{z}}, \hat{M}\}$ of commuting operators, where the
constant of motion $\hat{M}$ reads
%=================================================================
\bea \hat{M}=\hat{J}^{2}+\frac{2c_{1}}{1+\cos\theta}+
\frac{2c_{2}}{1-\cos\theta}.\label{7} \eea
%=================================================================
Here $\hat{J}^{2}$ is the square of the angular momentum,
$\hat{J_{z}}=s-i\partial/\partial\varphi$ its $z$-component and
$\hat{J_{z}}\psi=m\psi$.

After substitution the expression (\ref{6}) the variables in the
Schr\"{o}dinger equation are separated and we arrive at the
following system of coupled differential equations:
%===========================================================================
\bea
\frac{1}{\sin\theta}\frac{\partial}{\partial\theta}\left(\sin\theta\frac{\partial
Z}{\partial\theta}\right) +
\frac{1}{4\cos^2\frac{\theta}{2}}\left(\frac{\partial^{2}}{\partial\varphi^{2}}-4c_1\right)Z+
\nonumber \\ [3mm]
+\frac{1}{4\sin^2\frac{\theta}{2}}\left[\left(\frac{\partial}{\partial\varphi}+2is\right)^{2}
-4c_2\right]Z =-{\cal A}Z,\label{8}\\[3mm]
\frac{1}{r^{2}}
\frac{d}{dr}\left(r^{2}\frac{dR}{dr}\right)-\frac{\cal A}{r^{2}}R+
2\left(E+\frac{1}{r}\right)R=0, \label{9}\eea
%==============================================================================
where $\cal A$ is a separation constant in spherical coordinates.

The solution of (\ref{8}) is easily found to be
%====================================================================
\bea Z_{jm}^{(s)}(\theta, \varphi; \delta_{1}, \delta_{2}
)=N_{jm}(\delta_{1},
\delta_{2})\left(\cos\frac{\theta}{2}\right)^{m_1}
\left(\sin\frac{\theta}{2}\right)^{m_2}
P_{j-m_+}^{(m_2,m_1)}(\cos\theta) e^{i(m-s)\varphi}, \label{10}
\eea
%======================================================================
where $m_1=|m-s|+\delta_{1}=\sqrt{(m-s)^2+4c_1}$,
$m_2=|m+s|+\delta_{2}=\sqrt{(m+s)^2+4c_2}$, $m_+=(|m+s|+|m-s|)/2$
and $P_n^{(a,b)}$ denotes a Jacobi polynomial. The quantum numbers
$m$ and $j$ run through values: $m=-j,-j+1,\dots,j-1,j$ and
%=====================================================================
\bea j = \frac{|m+s|+|m-s|}{2}, \frac{|m+s|+|m-s|}{2}+1,\dots.
\nonumber \eea
%=====================================================================
The quantum numbers $j$, $m$ characterize the total momentum of
the system and its projection on the axis $z$. For the
(half)integer $s$ $j, m$ are (half)integers.

Furthermore, the separation constant $\cal A$ is quantized as
%======================================================================
\bea {\cal A} = \left(j+\frac{\delta_{1}+\delta_{2}}{2}\right)
\left(j+\frac{\delta_{1}+\delta_{2}}{2}+1\right). \label{11} \eea
%======================================================================
The normalization constant $N_{jm}(\delta_{1}, \delta_{2})$ in
(\ref{10}) is given (up to a phase factor) by
%=====================================================================
\bea N_{jm}(\delta_{1},
\delta_{2})=\frac{1}{2^{m_{+}}}\sqrt{\frac{(2j+\delta_{1}+\delta_{2}+1)(j-m_+)!
\Gamma(j+m_+
+\delta_{1}+\delta_{2}+1)}{2^{\delta_{1}+\delta_{2}+2}\pi\Gamma(j-m_-
+\delta_{1}+1) \Gamma(j+m_- +\delta_{2}+1)}}, \label{12} \eea
%=========================================================================
where $m_- = (|m+s|-|m-s|)/2$. The angular wavefunctions
$Z_{jm}^{(s)}$ [see Eq.(\ref{10})] are convenient to call the
ring-shaped monopole harmonics by analogy with the term "monopole
harmonics" studied by Tamm \cite{tamm}. These ring-shaped monopole
harmonics generalize the functions studied by Hartmann \cite{hart}
in the case $s=0, \delta_1=\delta_2$. Due to the connecting
formula \cite{erd}
%=========================================================================
\bea \left(\lambda +\frac{1}{2}\right)_{n}C_{n}^{\lambda}(x) =
(2\lambda)_{n}P_{n}^{(\lambda -\frac{1}{2},\lambda
-\frac{1}{2})}(x) \label{13} \eea
%=========================================================================
between the Jacobi polynomial $P_{n}^{(a,b)}$ and the Gegenbauer
polynomial $C_{n}^{\lambda}$, the case $s=0$,
$\delta_1=\delta_2=\delta$ yields
%========================================================================
\bea Z_{jm}^{(0)}(\theta, \varphi; \delta,
\delta)=2^{|m|+\delta}\Gamma\left(|m|+\delta
+\frac{1}{2}\right)\sqrt{\frac{(2j+2\delta+1)(j-|m|)!}{4\pi^2\Gamma(j+|m|+2\delta+1)}}
\nonumber \\ [3mm]
\left(\sin\theta\right)^{|m|+\delta}C_{j-|m|}^{|m|+\delta+\frac{1}{2}}(\cos\theta)
e^{im\varphi}, \label{14} \eea
%========================================================================
the result already obtained in Ref. \cite{KMP}. [In (\ref{13})
$(a)_n$ stands for a Pochhammer symbol.] The case $\delta=0$ (i.e.
$c_{1}=c_{2}=0$) can be treated by using the connecting formula
%========================================================================
\bea
P_{j}^{|m|}(x)=\frac{(-2)^{|m|}}{\sqrt{\pi}}\Gamma\left(|m|+\frac{1}{2}\right)
\left(1-x^{2}\right)^{\frac{|m|}{2}}C_{j-|m|}^{|m|+\frac{1}{2}}(x)
\label{15} \eea
%========================================================================
between the Gegenbauer polynomial $C_{n}^{\lambda}$ and the
associated Legandre function \cite{erd}. In fact for the
$\delta=0$, Eq.(\ref{14}) can be reduced to
%========================================================================
\bea Z_{jm}^{(0)}(\theta, \varphi; 0,
0)=\sqrt{\frac{(2j+1)(j-|m|)!}{4\pi (j+|m|)!}}
P_{j}^{|m|}(\cos\theta) e^{im\varphi}, \label{16} \eea
%========================================================================
an expression (up to a phase factor)that coincides with the usual
(surface) spherical harmonics $Y_{lm}(\theta, \varphi)$.

Let us go now to radial equation (\ref{9}). The introduction of
(\ref{11}) into the (\ref{9}) leads to
%========================================================================
\bea \frac{1}{r^{2}} \frac{d}{dr}\left(r^{2}\frac{dR}{dr}\right)-
\frac{1}{r^{2}}\left(j+\frac{\delta_{1}+\delta_{2}}{2}\right)
\left(j+\frac{\delta_{1}+\delta_{2}}{2}+1\right)R+
2\left(E+\frac{1}{r}\right)R=0, \label{17}\eea
%========================================================================
which is reminiscent of the radial equation for the hydrogen atom
except that the orbital quantum number $l$ is replaced here by
$j+(\delta_{1}+\delta_{2})/2$. The solution of (\ref{17}) for the
discrete spectrum is
%===============================================================
\bea R_{nj}^{(s)}(r)= C_{nj}(\delta_1, \delta_2)(2\varepsilon
r)^{j+\frac{\delta_1+ \delta_2}{2}}e^{-\varepsilon
r}F\left(-n+j+1; 2j+\delta_1+ \delta_2+2; 2\varepsilon r\right),
\label{18}\eea
%================================================================
where $n=|s|+1, |s|+2,\dots.$ In (\ref{18}), the normalization
factor $C_{nj}(\delta_1, \delta_2)$ reads
%================================================================
\bea C_{nj}(\delta_1, \delta_2)=
\frac{2\varepsilon^{2}}{\Gamma\left(2j+\delta_1+
\delta_2+1\right)}\sqrt{\frac{\Gamma\left(n+j+\delta_1+
\delta_2+1\right)}{(n-j-1)!}}\label{19}\eea
%=================================================================
and the parameter $\varepsilon$ is defined by
%=================================================================
\bea \varepsilon= \sqrt{-2E} = \frac{1}{n+\frac{\delta_1+
\delta_2}{2}}. \label{20}\eea
%=================================================================
The eigenvalues $E$ are then given by
%=================================================================
\bea E \equiv E_n^{(s)} = -\frac{1}{2\left(n+\frac{\delta_1+
\delta_2}{2}\right)^{2}}. \label{energy}\eea
%================================================================
In the limiting case $\delta_1= \delta_2 = 0$, we recover the
familiar results for charge-dyon bound system \cite{Z}.

\setcounter{equation}{0}

\section{Parabolic Basis}

Let us consider the generalized MIC-Kepler system in the parabolic
coordinates. In the parabolic coordinates $\xi,\eta \in [0,
\infty), \, \varphi \in [0, 2\pi)$, defined by the formulae
%==================================================================
\beq x = \sqrt{\xi \eta}\,\cos\varphi, \qquad y = \sqrt{\xi
\eta}\,\sin\varphi,\qquad z = \frac{1}{2}(\xi - \eta),
\label{parabolic}\eeq
%======================================================================
the differential elements of length and volume read
%==================================================================
\beq dl^2 = \frac{\xi + \eta}{4}\left(\frac{d\xi^2}{\xi} +
\frac{d\eta^2}{\eta}\right) + \xi \eta d\varphi^2, \qquad dV=
\frac{1}{4}(\xi + \eta)d\xi d\eta d\varphi, \label{pmetric} \eeq
%======================================================================
while the Laplace operator looks like
%====================================================================
\beq \Delta =\frac{4}{\xi + \eta}\left[\frac{\partial}{\partial
\xi} \left(\xi \frac{\partial}{\partial \xi}\right) +
\frac{\partial}{\partial \eta}\left(\eta \frac{\partial} {\partial
\eta}\right)\right] + \frac{1}{\xi \eta}
\frac{\partial^2}{\partial \varphi^2}. \label{laplasian} \eeq
%===================================================================
The substitution
%==================================================================
\beq \psi(\xi,\eta,\varphi) = \Phi_1(\xi)
\Phi_2(\eta)\,\frac{e^{i(m-s)\varphi}}{\sqrt{2\pi}}.
\label{parwave}\eeq
%======================================================================
separates the variables in the Schr\"{o}dinger equation and we
arrive at the following system of equations
%==================================================================
\begin{eqnarray}
\frac{d}{d \xi}\left(\xi \frac{d\Phi_1}{d \xi}\right) +
\left[\frac{E}{2}\xi - \frac{m_1^2}{4\xi} +
\frac{1}{2}\beta + \frac{1}{2}\right]\Phi_1 &=& 0, \label{eq1}\\
[3mm] \frac{d}{d \eta}\left(\eta \frac{d\Phi_2}{d \eta}\right) +
\left[\frac{E}{2}\eta - \frac{m_2^2}{4\eta}- \frac{1}{2}\beta +
\frac{1}{2}\right]\Phi_2 &=& 0, \label{eq2}
\end{eqnarray}
%======================================================================
where $\beta$ -- is the separation constant.

These equations are analogous with the equations of the hydrogen
atom in the parabolic coordinates \cite{landau}. Thus, we get
%==================================================================
\begin{eqnarray}
\psi_{n_1n_2m}^{(s)}(\xi,\eta,\varphi;\delta_1,\delta_2) =
\sqrt{2}\varepsilon^{2}\Phi_{n_1m_1}(\xi)
\Phi_{n_2m_2}(\eta)\,\frac{e^{i(m-s)\varphi}}{\sqrt{2\pi}},
\label{parwave1}
\end{eqnarray}
%======================================================================
where
%==================================================================
\begin{eqnarray}
\Phi_{n_im_i}(x) = \frac{1}{\Gamma(m_i+1)}
\sqrt{\frac{\Gamma(n_i+m_i+1)}{(n_i)!}}\,\,e^{-\frac{\varepsilon
x}{2}}\,\,(\varepsilon x)^{\frac{m_i}{2}}\,\,F(-n_i; m_i+1;
\varepsilon x). \label{parwave2}
\end{eqnarray}
%======================================================================
Here $n_1$ and $n_2$ are nonnegative integers
%==================================================================
\begin{eqnarray}
n_1 = -\frac{|m-s|+\delta_1+1}{2}+\frac{\beta+1}{2\varepsilon},
\qquad n_2 =
-\frac{|m+s|+\delta_2+1}{2}-\frac{\beta-1}{2\varepsilon}.
\label{parqnumb}
\end{eqnarray}
%======================================================================
From the last relations, taking into account (\ref{energy}), we
get that the parabolic quantum numbers $n_1$ and $n_2$ are
connected with the principal quantum number $n$ as follows
%==================================================================
\begin{eqnarray}
n= n_1 + n_2 + \frac{|m-s|+|m+s|}{2}+1. \label{parqnumb1}
\end{eqnarray}
%======================================================================
Excluding the energy $E$ from Eqs. (\ref{eq1}) and (\ref{eq2}), we
obtain the additional integral of motion
%==================================================================
\begin{eqnarray}
\hat{X} &=& \frac{2}{\xi + \eta}\left[\xi \frac{\partial}{\partial
\eta}\left(\eta \frac{\partial}{\partial \eta}\right) -
\eta\frac{\partial}{\partial \xi}\left(\xi
\frac{\partial}{\partial \xi}\right)\right] +
\frac{\xi-\eta}{2\xi\eta}\frac{\partial^{2}}{\partial\varphi^{2}}
+ is\frac{\xi^{2}+\eta^{2}}{\xi\eta
(\xi+\eta)}\frac{\partial}{\partial\varphi}- \nonumber \\
\\
&&-s^{2}\frac{\xi-\eta}{2\xi\eta} +
\frac{2c_{1}\eta}{\xi(\xi+\eta)} -
\frac{2c_{2}\xi}{\eta(\xi+\eta)} +
\frac{\xi-\eta}{\xi+\eta}\nonumber \label{int}
\end{eqnarray}
%======================================================================
with the eigenvalues
%==================================================================
\begin{eqnarray}
\beta =
\varepsilon\left(n_{1}-n_{2}+\frac{|m-s|-|m+s|+\delta_{1}-\delta_{2}}{2}\right)
\label{beta}
\end{eqnarray}
%======================================================================
and eigenfunctions
$\psi_{n_1n_2m}^{(s)}(\xi,\eta,\varphi;\delta_1,\delta_2)$.

In Cartesian coordinates, the operator $\hat{X}$ can be rewritten
as
%==================================================================
\begin{eqnarray}
\hat{X} &=& z\left(\frac{\partial^{2}}{\partial x^{2}} +
\frac{\partial^{2}}{\partial y^{2}}\right) -
x\frac{\partial^{2}}{\partial x \partial z} -
y\frac{\partial^{2}}{\partial y \partial z}  +
is\frac{r+z}{r(r-z)}\left(x\frac{\partial}{\partial
y}-y\frac{\partial}{\partial x}\right) - \nonumber \\
\\
&&- \frac{\partial}{\partial z}-s^{2}\frac{r+z}{r(r-z)} +
c_1\frac{r-z}{r(r+z)} - c_2\frac{r+z}{r(r-z)} + \frac{z}{r},
\nonumber \label{intc}
\end{eqnarray}
%======================================================================
so that it immediately follows that $\hat{X}$ is connected to the
$z$-component $\hat{I_z}$ of the analog of the Runge-Lenz vector
(\ref{3}) via
%==================================================================
\begin{eqnarray}
\hat{X} =\hat{I_z} + c_1\frac{r-z}{r(r+z)} - c_2\frac{r+z}{r(r-z)}
\label{intc1}
\end{eqnarray}
%======================================================================
and coincides with $\hat{I_z}$ when $c_1=c_2=0$.

Thus we have solved the spectral problem in spherical
%==================================================================
\begin{eqnarray}
\hat{\cal H}\psi =E\psi, \qquad {\hat
M}\psi=\left(j+\frac{\delta_1+\delta_2}{2}\right)
\left(j+\frac{\delta_1+\delta_2}{2}+1\right)\psi, \qquad
\hat{J_z}\psi=m\psi \label{spectr1}
\end{eqnarray}
%======================================================================
and in parabolic coordinates
%==================================================================
\begin{eqnarray}
\hat{\cal H}\psi =E\psi, \qquad {\hat X}\psi=\beta\psi, \qquad
\hat{J_z}\psi=m\psi, \label{spectr2}
\end{eqnarray}
%======================================================================
where $\hat{\cal H}$, $\hat{J_z}$, ${\hat M}$ and ${\hat X}$ are
defined by the expressions (\ref{1}), (\ref{3}), (\ref{7}) and
(\ref{intc1}).

It is mentioned that all the formulae obtained for $s=0$ yield the
corresponding formulae for the generalized Kepler-Coulomb system
\cite{KMP}.

\vspace{5mm}

{\large Acknowledgements.} The work is carried out with the
support of ANSEF No: PS81 grant.

\end{document}